\documentclass[sigplan,nonacm]{acmart}

\AtBeginDocument{%
  }

\setcopyright{acmlicensed}
\copyrightyear{2018}
\acmYear{2018}
\acmDOI{XXXXXXX.XXXXXXX}
\acmConference[Conference acronym 'XX]{Make sure to enter the correct
  conference title from your rights confirmation email}{June 03--05,
  2018}{Woodstock, NY}
\acmISBN{978-1-4503-XXXX-X/2018/06}

\usepackage{booktabs}
\usepackage{multirow}
\usepackage{makecell}
\usepackage{subfigure}
\usepackage[ruled,linesnumbered]{algorithm2e}
\newcommand{\phb}[1]{\vspace{.4em} \noindent\textbf{#1}\hspace{.5em}} 

\begin{document}

\title{A Flexible Programmable Pipeline Parallelism Framework 
for Efficient DNN Training}

\author[Lijuan Jiang et. al.]{Lijuan Jiang$^1$, Xingjian Qian$^2$, Zhenxiang Ma$^3$, Zan Zong$^{4}$, Hengjie Li$^1$, Chao Yang$^{5}$, Jidong Zhai$^{4}$}
% \thanks{*Corresponding author.}
\affiliation{
    \institution{$^1$ Shanghai AI Lab,  $^2$ Zhejiang University, $^3$ Shanghai Jiao Tong University \\ $^4$ Tsinghua University, $^5$ Peking University}
    \country{}
}
\begin{abstract}
Pipeline parallelism is an essential distributed parallelism method. Increasingly complex and diverse DNN models necessitate
meticulously customized pipeline schedules for performance. However, existing practices typically rely on predefined schedules, each
with strengths, but fail to adapt automatically to the emerging model architectures. Exploring novel high-efficiency schedules is daunting due to the enormous and varying schedule space.
Besides, manually implementing schedules can be challenging due to the onerous coding burdens and constantly changing needs. Unfortunately, existing frameworks have limitations in automated schedule exploration and lack flexibility and controllability.

This paper presents FlexPipe, a programmable pipeline parallelism framework with enhanced productivity, program- mability, debuggability, and ease of tuning. FlexPipe has two main components: a succinct domain-specific language (DSL) and an automated scheduler. FlexPipe enables automated schedule exploration for various parallel scenarios within a broad spectrum of schedule types at a small search cost. Besides, users can swiftly develop and customize schedules using the FlexPipe DSL, which embodies flexible controllability in the pipeline order of micro-batch computations over stages. It also provides convenient mechanisms to include new operations in schedules to meet changing demands. Our evaluation results demonstrate that FlexPipe achieves up to 2.28$\times$ performance speedup compared to the popular large-scale parallel framework Megtron-LM, and gains up to 1.49$\times$ performance speedup compared to the state-of-the-art automated pipeline parallelism framework.
\end{abstract}



\maketitle

\section{Introduction}
Large-scale DNN models 
% based on deep transformer layers and massive datasets 
have gained prominent performance in various areas such as natural language processing~\cite{achiam2023gpt,dubey2024llama}, computer vision~\cite{dosovitskiy2020image}, text-to-video~\cite{liu2024sora}, etc. Besides, model sizes~\cite{radford2018improving,radford2019language,brown2020language,achiam2023gpt} increase by leaps and bounds. 
Distributed parallelism~\cite{jiang2024megascale} has been the fundamental method for training large-scale models due to the long training time and massive memory consumption brought about by the ever-increasing model sizes. Furthermore, pipeline parallelism is generally used as an indispensable model parallelism method to scale up to larger models across servers~\cite{narayanan2021efficient}.

Pipeline parallelism~\cite{huang2019gpipe} shards the model at the layer level 
% and relies
% on lightweight peer-to-peer communication to transfer intermediate data, making it highly scalable.
% and indispensable in large-scale distributed training.
% Consecutive model layers are primarily grouped 
into multiple stages that are placed on different devices. 
% To further relieve the device idling time caused by the activation communication, 
Next, a mini-batch of training samples is split 
into smaller micro-batches, the execution of which over stages is pipelined to allow devices to work simultaneously. 
% However, two main factors hinder the efficiency of pipeline parallelism, i.e., bubbles and memory consumption. Bubbles refer to the device idle time due to data dependency between the boundary layers of stages across different devices. Memory consumption comes from model parameters and activation memory, etc. 
% Notably, considerable communication can also become a key factor that limits efficiency.
Both the spatial stage placement and the temporal schedule that decides the pipelined execution order of micro-batch computations are critical to the efficiency of pipeline parallelism in terms of device idle time (also referred to as bubbles) and memory consumption. %~\cite{lin2024tessel}

Extensive research~\cite{fan2021dapple,narayanan2021efficient,li2021chimera,qi2023zero} has proposed effective stage placement strategies and schedules.
However, current approaches usually rely on predefined schedules, each with strengths, but fail to adapt automatically to the emerging model architectures.
% It is difficult for new scenarios, where advancements in model architectures 
% are made (\S~\ref{sec:background}), to benefit from existing methods. 
% Existing methods mainly rely on predefined schedules, which can introduce additional bubbles and thus hurt performance. 
Moreover, exploring novel efficient schedules is daunting due to the following two aspects.

% First, finding efficient schedules can be time-consuming.
First, the schedule space is enormous.
 % and prone to errors
It typically requires experts to handcraft ingenious schedules, managing hundreds or even thousands of micro-batches and their intricate dependencies while allowing for multiple in-flight micro-batches for pipeline efficiency.
% allowing for multiple in-flight micro-batches concerning pipeline efficiency and memory consumption. However, the schedule space is enormous as it often involves . 
% to enhance device utilization
% the schedule space is large. 
% Suppose the stage placement is specified, 
% , one needs to explore the temporal schedules.
% Different schedules for micro-batch computations exhibit varying memory costs and bubbles. 
% It typically requires experts to meticulously arrange complex schedules by hand, allowing for multiple in-flight micro-batches to enhance device utilization. 
% However, this approach can be time-consuming and prone to errors, as it often involves managing hundreds or even thousands of micro-batches along with their intricate dependencies.
Various stage placement strategies further complicate the schedule space. Multiple stages are placed on the same device, and the execution order of micro-batches over different stages needs to be determined. 
% The second challenge stems from the sophisticated performance characteristics of various pipeline approaches.
In addition, the performance bottlenecks of schedules depend not only on the method and the underlying hardware but also on the model's inputs. 
As a result, the explored schedule running on the same machine may not be usable or fail to achieve optimal performance given different model inputs. 
For example, Interleaved 1F1B demonstrates superior performance over 1F1B at small batch sizes. Otherwise, the latter schedule achieves better performance since the influence of bubbles is insignificant, while the former incurs more communications.
% For example, bidirectional pipeline parallelism~\cite{li2021chimera,wu2024bitpipe} that maintains a more balanced memory consumption appears to be more suitable for transformer-based models with large embedding layers, while others may encounter the out-of-memory (OOM) issue. 
% Besides, the hyperparameters, such as the number of partitions
% stages and the micro-batch size also have a non-ignorable influence on the performance.

Second, manually implementing schedules for various distributed scenarios exposes significant challenges: 1)\textit{Productivity}. Manually customizing and developing schedules requires onerous coding burdens and expertise. The implementation of 1F1B and Interleaved 1F1B contains around 0.7k and 1.4k code lines, respectively, in the popular framework Megatron-LM~\cite{megatronlm-github}. 2)\textit{Programmability}. Developers must maintain tangled communications and data dependencies concerning different micro-batches, computation types, and even stages in distributed environments. In addition, diverse model architectures may constantly introduce new operations in schedules. For instance, synchronization is required to exchange the output states of submodules corresponding to different modalities in multi-modal models~\cite{radford2021learning}. 3)\textit{Debuggability}. Manual implementations of schedules are prone to errors and lack intermediate debugging information, making the debugging process very time-consuming and unbearable.

Existing effective frameworks~\cite{lin2024tessel} enable automated schedule exploration when the stage placement strategy is specified. However, it shows deficiencies in the diversity of searchable schedules. For example, it cannot support the circular stage placement strategy, with which multiple types of efficient schedules are designed for different parallel scenarios~\cite{narayanan2021efficient,liu2023hanayo,lamy2023breadth,qi2024pipeline}. 
Besides, the search cost is intolerable when the hardware resources are extensive.
Compiler approaches~\cite{tang2025koala} propose a domain-specific language (DSL) to improve productivity in developing schedules.
% However, the DSL is based on a sophisticated syntax with a two-layer structure, including a computational graph DSL to describe schedule templates and an executable DSL to elaborate on the specific runtime details, respectively. 
However, implementing 1F1B still requires 0.2k code lines due to the sophisticated two-layer structured syntax. In addition, the compiler approach can neither enable automated exploration of novel schedules nor support new operations in schedules for various model architectures.
% Compiler approaches~\cite{xu2021gspmd,chen2024slapo} typically support only one or two fixed types of pipeline parallelism methods and struggle to flexibly control or tune schedules with existing domain-specific languages (DSL). 

In this paper, we present FlexPipe, a flexible programmable pipeline parallelism framework that circumvents limitations in existing frameworks.
FlexPipe includes a succinct DSL to express pipeline schedules, enabling automated schedule exploration within a broad spectrum of schedule types at small overheads. With FlexPipe, implementing existing mainstream schedules only requires a few lines of code. Besides, FlexPipe provides mechanisms to control the pipeline order of micro-batch computations over stages flexiblely and to support new operations to customize schedules swiftly.

In FlexPipe, we regard the scheduling of a micro-batch computation (i.e., forward or backward pass) as a scheduling step.
We observe that the key to exploring and developing schedules is the flexible controllability in the scheduling order of micro-batch computations.
We also observe that, suppose the stage placement strategy is specified, the schedule can be broken down into a series of scheduling steps. The forward or backward passes are constantly selected to be scheduled from the micro-batch computations with resolved dependencies on each device.
We can control the schedule by determining the scheduling priorities, which decide the order in which the dependency-free micro-batch computations are selected for scheduling.

Two types of scheduling priorities need to be determined: 1)the priority concerning computation types (we call \textbf{computation type traversal priority}), i.e., whether to schedule preferentially forward or backward passes; 2)the priority concerning stages (we call \textbf{stage traversal priority}) that decides the micro-batch computations of which stage are preferentially to be scheduled if multiple stages are placed on the same device. Furthermore, based on the concepts of scheduling priorities, we observe that efficient schedules generally employ the same scheduling priorities in different scheduling steps. 
% These observations can largely enhance the flexibility and controllability of the complicated schedule process and mitigate the schedule exploration overhead.

% Based on the above observations, we introduce FlexPipe, which translates these insights into a flexible programmable pipeline parallelism framework
% for automatic schedule exploration and circumvents limitations in existing frameworks. On top of FlexPipe, we provide a user-friendly DSL to succinctly express most existing mainstream schedules and design new schedules without worrying about the complex dependencies. In addition, FlexPipe includes an automated
% scheduler that incorporates a computation schedule space representation (CSSR) and an actor-aware schedule mechanism. An auto-tuner is designed to tune stage placements and schedules for different model inputs and hardware resources.
FlexPipe is built upon the above essential observations.
Besides the FlexPipe DSL, FlexPipe also includes an automated scheduler that interacts with the DSL 
to arrange schedules automatically, and an auto-tuner that finds the best configurations for various model inputs.
FlexPipe facilitates the schedule exploration with enhanced productivity, programmability, debuggability, and ease of tuning.
Compared with state-of-the-art methods, our experiments demonstrate that FlexPipe achieves up to 2.28$\times$ performance speedup on training language models with large embedding layers and up to 1.29$\times$ performance speedup in multimodal models. Overall, this work makes the following contributions.

\begin{itemize}
    \item It introduces a succinct DSL that allows flexible controllability in the pipeline order of micro-batch computations to express various schedules.
    %[基于schedule primitive的方式，增加了灵活性和可控执性，]自动搜索schedule 扩展搜索结果，并将搜索时间限制在数秒内（搜索时间如果跟其它并行方法联合时非常必要）
    \item It enables automated schedule exploration with diverse searchable schedule types at small overheads for various stage placement strategies.
    \item It provides programmable mechanisms to customize schedules for the emerging model architectures without worrying about tanglesome data dependencies, communications, etc.
    \item It proposes FlexPipe, an end-to-end system, which instantiates explored schedules for efficient runtime execution. 
    Besides, it can also tune the best configuration of schedule types and hyperparameters for various model inputs and hardware resources. %通过预定义的基本stage placement，参数化预定义schedule，可以tune，用户可以基于可编程接口，设计新型流水线而不用在意通信、依赖关系等问题 
    % \item It evaluates FlexPipe in several new scenarios and compares its performance with state-of-the-art works. Results show that ...%在新场景中的一些加速效果
\end{itemize}

\section{Background and Motivation}\label{sec:background}
% In this section, we first introduce common concepts of pipeline parallelism, followed by motivating examples to illustrate the challenges of schedule exploration.
% We use data parallelism~\cite{fan2021dapple} to scale up the model and leverage tensor parallelism~\cite{shoeybi2019megatron} to optimize stage placement strategies. 
% % Other parallelization approaches, such as 3D parallelism
% Other parallel methods~\cite{narayanan2021efficient} are orthogonal to our approaches, but are beyond the scope of our discussion. 

\subsection{Pipeline Parallelism.}
% Pipeline parallelism methods arrange schedules that generally allow multiple in-flight micro-batches for reduced bubbles and peak memory consumption.
% Pipeline parallelism typically consists of three phases: the warm-up, steady, and cool-down phases~\cite{narayanan2019pipedream}. Enough micro-batches of computations are launched in the warm-up phase. In the steady phase, forward and backward passes are calculated regularly. In the cool-down phase, the left computations are performed. 
% Besides, peer-to-peer (P2P) communications are required to transfer activations and gradients during the forward and backward propagation, respectively.
Prior research has proposed effective stage placement strategies.
% in conjunction with carefully arranged . 
Figure~\ref{fig-placements}(i) shows that 1F1B uses the one-to-one stage placement strategy. Interleaved 1F1B~\cite{narayanan2021efficient}, 
Hanayo~\cite{liu2023hanayo} and Chimera~\cite{li2021chimera} propose the circular, V-shape, and bidirectional stage placement strategies, respectively,  where bubbles are reduced due to the smaller computational time of a single micro-batch, or more devices working simultaneously.
Subsequent research further evolves based on the above stage placement strategies.
% Breadth-first pipeline parallelism~\cite{lamy2023breadth} primarily schedules all micro-batches of forward passes based on the circular stage placement strategy for the combination of pipeline and data parallelism. 
BitPipe~\cite{wu2024bitpipe} fuses interleaved pipelines with bidirectional pipelines.
Furthermore, ZB-H1~\cite{qi2023zero} splits the gradient computation and fills the weight gradient computations in bubbles of 1F1B to improve efficiency.

% Different pipeline parallelism methods exhibit complicated performance, and none of the methods has the best performance under all scenarios. 
% and auto-tuning for different inputs. 
% In FlexPipe, we support all methods mentioned above.
% In addition, we design a ``stash-and-insert'' algorithm to leverage the gradient separation technique that automatically detects and fills in bubbles for different schedule types.

\subsection{Motivation}
% Diverse model architectures are continuously proposed for performance improvement in different scenarios. However, 
Diverse model architectures pose challenges to existing predefined pipeline schedules.

\begin{figure}[t]
\centerline{\includegraphics[width=\linewidth]{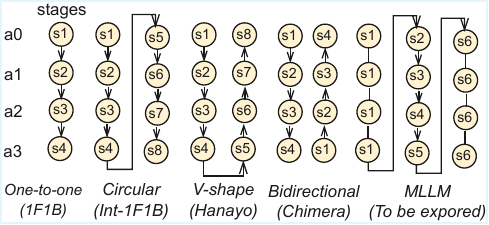}}
\caption{Effective stage placement strategies. A0-a4 represent devices, and s1-s8 represent stages.}
\label{fig-placements}
\end{figure}

\begin{figure}[t]
\centerline{\includegraphics[width=1.0\linewidth]{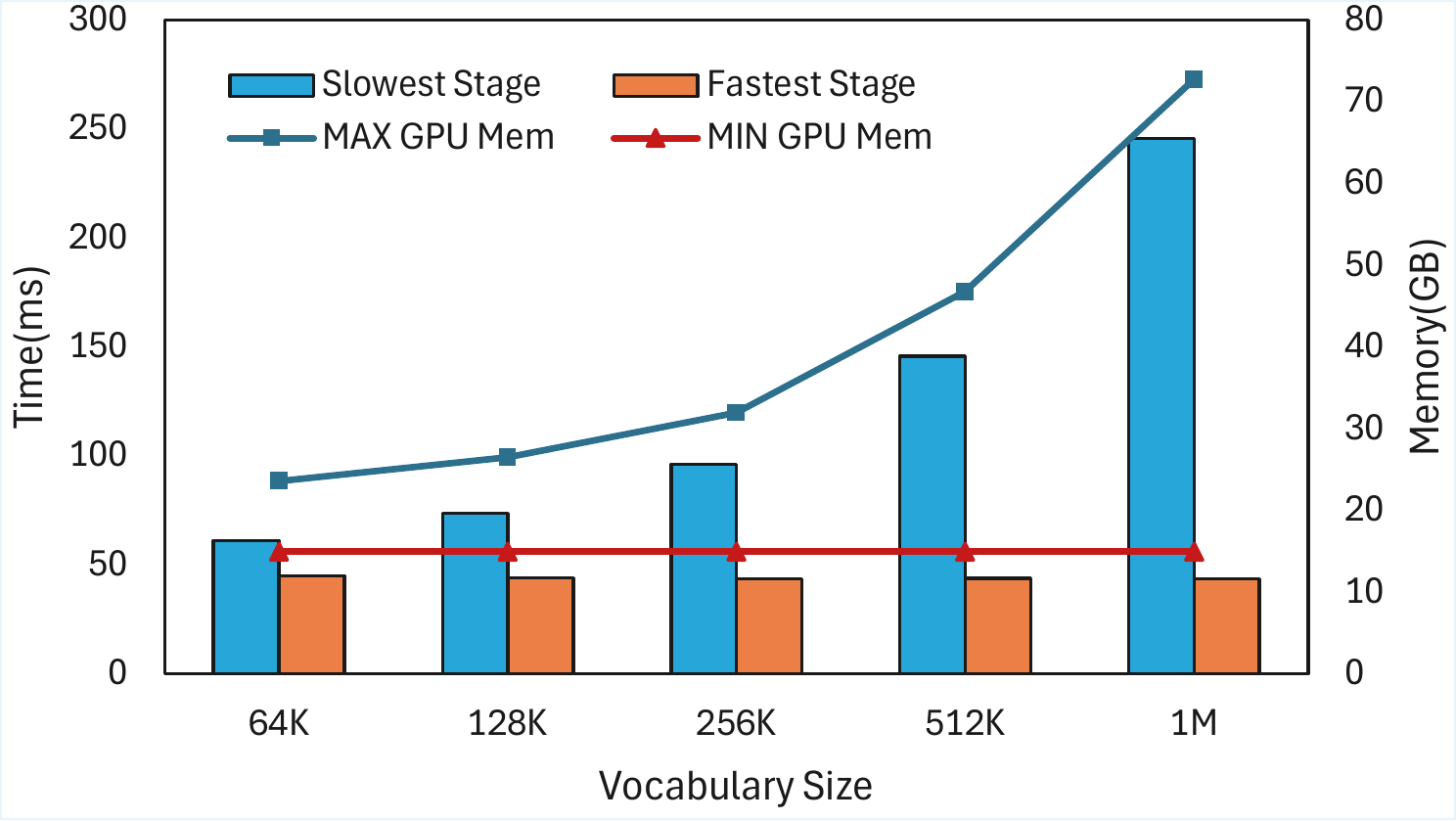}}
\caption{Imbalanced workloads of a 5B GPT model with varying vocabulary sizes.}
\label{fig-imbalanced}
\end{figure}

\phb{Schedule exploration given various stage placement strategies.}
% The widely applied transformer-stacked models are generally split evenly considering the transformer layers using the stage placements mentioned above. However, this may incur performance degradation in 
Exploring efficient schedules for different stage placement strategies is necessary but prohibitive due to the enormous schedule space. For instance, large embedding layers~\cite{zheng2021allocating,team2024gemma} in multilingual models are introduced to cover the vocabulary of multiple languages. However, using the existing stage placement strategies, the performance degrades due to the imbalanced workloads of the stages with and without embedding layers.
Figure~\ref{fig-imbalanced} profiles a GPT model to compare the imbalanced computations and memory consumption, which become increasingly imbalanced with the increasing vocabulary size. The slowest stage is 5.63$\times$ slower than the fastest for the GPT model with the 1M vocabulary size, and costs 4.87$\times$ more memory.
New stage placement strategies (``MLLM'' in Figure~\ref{fig-placements}) that distribute the embedding layers over multiple devices to balance memory and computations are proposed. However, extra bubbles are produced due to data dependencies when the specified stage placement strategy is applied to the predefined schedules. 
% Although Lin et al.~\cite{lin2024tessel} enables automated schedule exploration, more active micro-batches are launched at the start of the schedule to avoid bubbles in terms of data dependency in the steady phase, which in turn causes more bubbles in the warm-up and cool-down phases. Besides, only one schedule type is searched for the new stage placement strategy, but with large search overheads, e.g, above twenty hours on eight devices.

% \begin{figure}[t]
% \centerline{\includegraphics[width=\linewidth]{figs/1f1b_imbalanced.png}}
% \caption{Imbalanced pipeline workloads caused by large embedding layers under the 1F1B strategy.}
% \label{fig-1f1bimbal}
% \end{figure}

% \begin{figure}[t]
% \centerline{\includegraphics[width=\linewidth]{figs/perfdiff.png}}
% \caption{Performance comparison between several existing methods of different modalities.}
% \label{fig-perfdiff}
% \end{figure}

\begin{figure}[t]
\centerline{\includegraphics[width=1.0\linewidth]{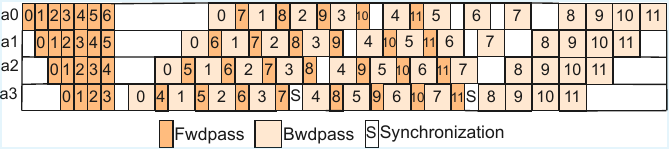}}
\caption{DistMM-Pipe.}
\label{fig-mov1f1b}
\end{figure}

\phb{Schedule customization for diverse model architectures.}
% \phb{Auto-tuning for different model configurations}
There is a demand for customizing schedules for various model architectures, which arrange micro-batch computations differently from existing predefined schedules. Besides, new operations may be required.
% supported, including the forward/backward pass, and activation/gradient communications that are supported by general pipeline schedules.
For instance, DistMM-Pipe~\cite{huang2024distmm} is designed for 
multi-modal models~\cite{driess2023palm,yu2022coca} that contain multiple submodules to process different modalities such as image and text.
% the gap between isolated data modalities in the real world, . 
DistMM-Pipe distributes different submodules onto different devices for parallel computation.
Since submodules need to synchronize the output states, DistMM-Pipe launches more forward passes in the warm-up phase and adds a synchronization operation every few micro-batches to avoid frequent communications, as shown in Figure~\ref{fig-mov1f1b}. The synchronization of submodules is supported as a new operation not included in existing schedules.
In addition, further analysis shows that synchronization costs around 17\%. Asynchronous communications can be leveraged to optimize bubbles. We can also employ mixed schedule types for different submodules since submodules are configured with varying model configurations and may be fit for different schedule types. However, efficient methods to flexibly tune schedules and support new operations are lacking.


\section{Overview of FlexPipe}
\begin{figure}[t]
\centerline{\includegraphics[width=1.0\linewidth]{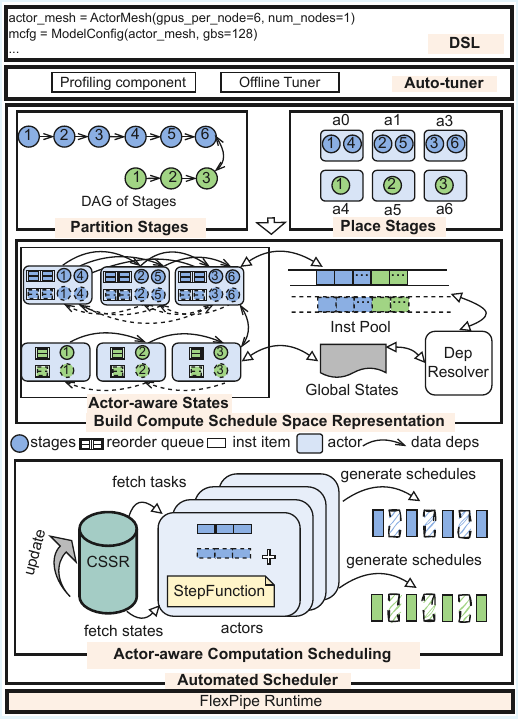}}
\caption{FlexPipe Overview.}
\label{fig-framework}
\end{figure}

We present the pipeline parallelism framework FlexPipe to screen users from the complexities of exploring and customizing efficient schedules for various parallel scenarios.
This section provides an overview of FlexPipe,
% FlexPipe performs automated search and tuning for efficient schedules given stage placement strategies.
as shown in Figure~\ref{fig-framework}, which consists of four components: a DSL, an automated scheduler, an auto-tuner, and a runtime.
% In FlexPipe, we call each device an actor.

FlexPipe DSL (\S~\ref{sec:dsl}) offers a simple API to express various schedules.
% At the bottom, the scheduler generates execution plans comprised of computation and communication tasks.
Users only need to specify the corresponding parameters in the API for different schedule types. 
% The main parameters can be set from the built-in values listed in Table~\ref{tab:compprimitives} to describe stage placement strategies and priorities controlling the pipeline order of micro-batch computations.
To include new operations in schedules, users only need to write a function for the new operation.
% for rapidly evolving model architectures. 
All the other complex scheduling is done by the framework. Furthermore, we provide more mechanisms for users to precisely control the scheduling of micro-batch computations as detailed in Section~\ref{sec:dsl-schd}.
% Based on the composition and operation mechanism of the scheduler, we abstract predefined values (listed in Table~\ref{tab:compprimitives}) in DSL for automated schedule arrangement and tuning in new scenarios.

The scheduler (\S~\ref{sec:lib}) arranges schedules automatically.
% which mainly consists of the computation schedule space representation (CSSR)
% and actor-aware schedule. 
It is worth noting that FlexPipe uses actors as the internal representation for devices.
The scheduler primarily constructs the intermediate representation of the computation schedule space (CSSR), inspired by conventional CPU instruction pipeline problems~\cite{crawford1990execution}.
CSSR leverages a set of instructions to represent various computations and communications.
Next, the actor-aware schedule is performed, which interacts with CSSR to automatically generate schedules represented as sequences of instructions according to the scheduling information specified in the DSL.
Besides, optimization passes such as gradient separation and asynchronous communications are included to optimize bubbles.
% It includes several components shown in Figure~\ref{fig-framework}. 
% Stage placement description places the stages on the actors. Then, the  is built. The schedule primitives control how the micro-batch computations are scheduled in order. The pipeline optimization passes aim at bubble reduction by leveraging the fine-grained gradient scheduling, memory optimizations including recomputation and CPU offloading, and asynchronous communication optimizations.

The auto-tuner (\S~\ref{sec:autotune}) searches and tunes efficient schedules under different model inputs and hardware resources. The auto-tuner consists of a profiling component and an offline tuner. The profiling component collects the timing metrics for computations and communications. Besides, the collected data is stored for future analysis, which avoids repeated data collection. The offline tuner finds the most suitable schedules enumerated in the schedule space.

% The workflow of FlexPipe is as follows. Users primarily provide new stage placement strategies. The tuner accesses the combinations of predefined values and drives the scheduler to generate different schedules. Besides, the schedule is further lowered into an acyclic task graph, which is used for the performance estimator to predict
% the performance. The tuner finally ranks the generated schedule according to the predicted performance.
% Finally, tasks in the execution plan are mapped to functions to run on accelerators.

\section{The FlexPipe DSL} \label{sec:dsl}
% FlexPipe's data representation consists of stage sets and actormesh, stage, and actor data.
% The FlexPipe DSL is embedded in Python. 
% The DSL allows users to control the schedule type.
% Additionally, the user can invoke the autotuner to automatically find the best-performing schedule for the given model definition and the underlying
% hardware platform.
The FlexPipe DSL includes constructs that express stages and a scheduling language to compose pipeline schedules.
The FlexPipe DSL is embedded in Python.
Users can specify schedules manually. Additionally, the auto-tuner can be invoked to find efficient schedules automatically.

\subsection{Data Model}\label{sec:dsl-data}
\begin{figure}[h]
\centerline{\includegraphics[width=\linewidth]{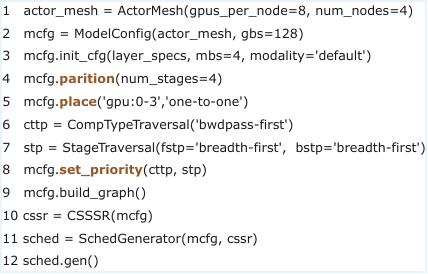}}
\caption{1F1B using the FlexPipe DSL.}
\label{fig-1f1bexp}
\end{figure}

% Users first define the model definition and actor topology mesh as inputs. Besides, the users can set the stage placements, schedule priority, and the number of in-flight micro-batches to control the schedule types as shown in Figure~\ref{fig-dsl}.
The primary constructs concerning stages consist of Stage, Actor, StageSet, and ActorMesh. 
Stage and StageSet describe the stage data and the DAG of stages, respectively.
Actor and ActorMesh express the device data and the device topology, respectively.
These constructs are operated through the construct ModelConfig. Users first define the device topology and model configuration (Lines 1-3 of Figure~\ref{fig-1f1bexp}).
Besides, users can set the configuration of different submodules by invoking the interface \textit{init\_cfg} multiple times with various settings of parameter ``modality'' (Line 3 of Figure~\ref{fig-1f1bexp}).

\begin{figure}[t]
\centerline{\includegraphics[width=1.03\linewidth]{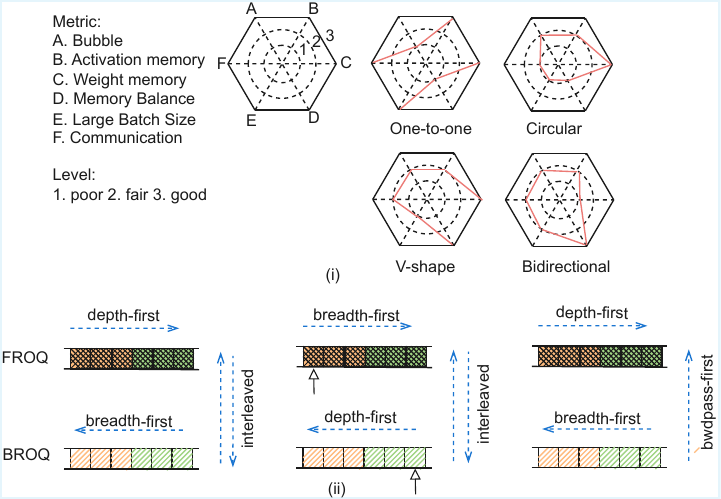}}
\caption{(i)Characteristics of each stage placement strategy. (ii)Illustrations on the traversal order of instruction items in reorder queues.}
\label{fig-metric}
\end{figure}

FlexPipe uses the method \textit{partition} and \textit{place} (Lines 4-5 of Figure~\ref{fig-1f1bexp}) to split model layers into stages and build mappings from stages to actors. By default, we evenly distribute the transformer layers into each stage.
Besides, we compare the strengths and weaknesses of different stage placement strategies in Figure~\ref{fig-metric}, and find that each stage placement strategy has strengths and weaknesses so that no single approach can outperform the other methods in all metrics. Hence, we have built in the one-to-one, circular, V-shape, and bidirectional stage placement strategies. In addition, a stage can be labelled as a shared stage among several actors. 
% When the shared Stage is detected in the scheduling process, a synchronization instruction is inserted automatically. 
A user must rewrite the \textit{partition} method when the default layer split rule fails to meet the demand.
% Besides, users can combine the predefined stage placement strategies. For instance, we can use \textit{VShapePlacement} and \textit{BiDirectionPlacement} together.
The partitioned stages are organized as a DAG (Line 9 of Figure~\ref{fig-1f1bexp}). 
% The data dependency settings 
% due to synchronization among submodules 
% are detailed in the following section.

\phb{FlexPipe's Instructions}
FlexPipe represents pipeline schedules as sequences of instructions mapped to the corresponding operations at runtime. Each instruction is featured with (stage ID, micro-batch ID).
Instructions in FlexPipe can be classified as (i) computations, including forward and backward pass, and the gradient calculation of weights and inputs,
% and the ''overlapped forward and backward pass'' specified in DualPipe, 
and (ii) cross-rank communications, including P2P send and receive for activations and gradients, and synchronizations for multimodal models.
The instructions supported by FlexPipe are listed in Table~\ref{tab:insts}.
Besides, FLexPipe supports new operations registered as new instructions, which will be
detailed in the following section.

\begin{table}[h]
    \centering
    \caption{Instructions of FlexPipe for various schedules.}
    \resizebox{\linewidth}{!}{
	\renewcommand\arraystretch{1.1}
    \begin{tabular}{cc}
    \hline
      \textbf{Computation} & \makecell[c]{FwdPass, BwdPass, \\ CompWeightGrad, CompInputGrad} \\
      \hline
      \textbf{Communication} & \makecell[c]{SendAct, SendGrad, RecvAct, RecvGrad, \\ SyncWithAllGather, SyncWithGather} \\
      
    \hline
    \end{tabular}
    }
    \label{tab:insts}
\end{table}

% \begin{table}[h]
%     \centering
%     \caption{Instructions of FlexPipe including all common computations and communications for pipeline schedules.}
%     \resizebox{\linewidth}{!}{
%     % \tabcolsep=1cm
% 	\renewcommand\arraystretch{3}
%     \begin{tabular}{|c|c|}
%     \hline
%        Computation &  Fwdpass, Bwdpass, CompWeightGrads, CompInputGrads, OverlappedFwdBwdpass \\
%       \hline
%       Communication & SendAct, SendGrad, RecvAct, RecvGrad, AllGather, Gather \\
%       \hline
%     \end{tabular}
%     }
%     \label{tab:instructions}
% \end{table}

\subsection{Scheduling Language}\label{sec:dsl-schd}
\begin{figure}[t]
\centerline{\includegraphics[width=1.0\linewidth]{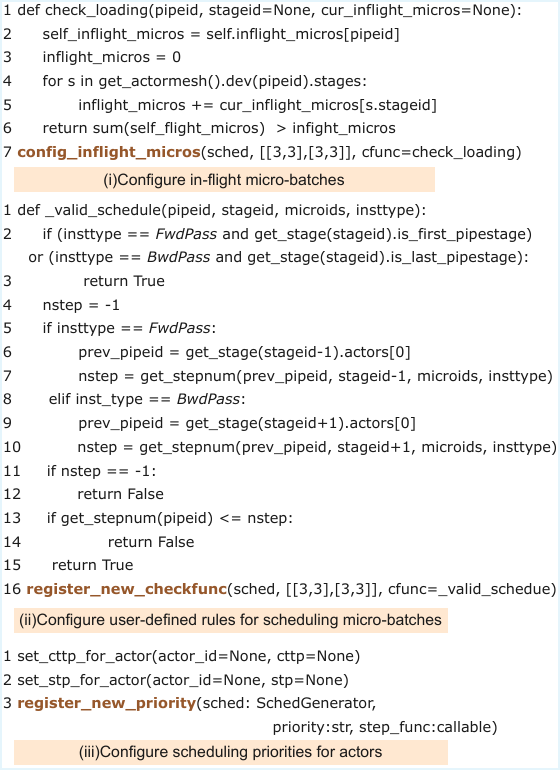}}
\caption{DSL Examples of controlling micro-batch orders.}
\label{fig-dslmicros}
\end{figure}

Users can specify pipeline schedules using FlexPipe's scheduling language. 
At the very least, users can specify schedule types by setting two parameters corresponding to two scheduling priorities: the computation type traversal priority and the stage traversal priority.
Table~\ref{tab:compprimitives} lists the built-in values for the two priorities.
Lines 6-8 of Figure~\ref{fig-1f1bexp} illustrate how the priorities are set. \textit{Set\_priority} sets the same priorities for actors assigned to a modality. 
% Users can also set 
Different priorities can be set for actors using \textit{set\_cttp\_for\_actor} and \textit{set\_stp\_for\_actor} (Line1,2 of Figure~\ref{fig-dslmicros}(iii)) for the two priorities, respectively.
We detail how we abstract the built-in values for the priorities and the scheduling mechanisms based on the priorities in \S~\ref{sec:lib}.

Moreover, the constructs CSSR and SchedGenerator (Line 10,11 of Figure~\ref{fig-1f1bexp}) describe the two main components of the automated scheduler (\S~\ref{sec:lib}). A user can control the scheduling of micro-batch computations through the methods and programmable mechanisms in terms of the two constructs.

% \subsubsection{Programmable Mechanisms}

\begin{figure}[b]
\centerline{\includegraphics[width=1.0\linewidth]{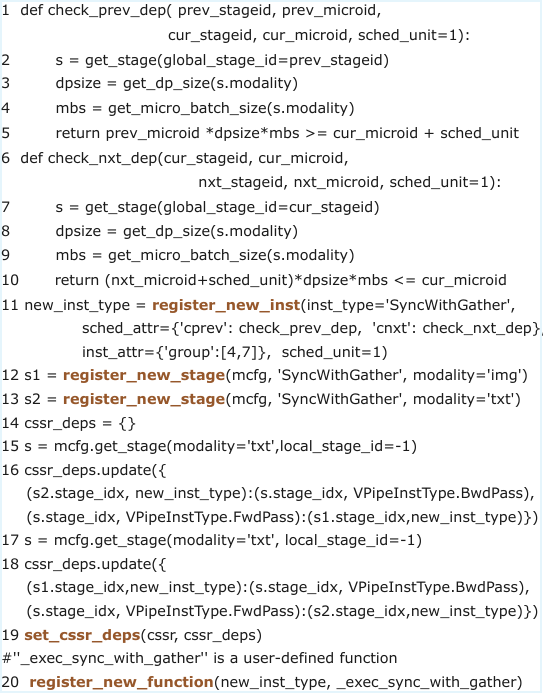}}
\caption{A multimodal example for registering instructions.}
\label{fig-dslmultimodal}
\end{figure}

\phb{Registration of new instructions.}
Users can support new operations that FlexPipe has not included yet. Figure~\ref{fig-dslmultimodal} presents the registration of synchronization for multimodal models to illustrate how to support and bind new operations with instructions for scheduling. 
First, \textit{register\_new\_inst} registers a new instruction and returns the corresponding instruction type (Line 11 of Figure~\ref{fig-dslmultimodal}). Users can configure the rules to schedule the registered instructions by specifying the attribute ``sched\_attr''. Lines 1-5 of Figure~\ref{fig-dslmultimodal} specify how many prior dependent instructions must have been scheduled before scheduling the registered instruction with the micro-batch ID ``cur\_microid''. Lines 6-10 of Figure~\ref{fig-dslmultimodal} specify when to schedule the subsequent dependent instructions with the micro-batch ID ``nxt\_microid''. The scheduling unit ``sched\_unit'' decides how many instructions are scheduled at a time. The scheduler (\S~\ref{sec:lib}) calls the ``check\_prev\_dep'' and ``check\_nxt\_dep'' automatically to resolve data dependencies during the scheduling process. Users can also configure the attributes ``inst\_attr'' for the corresponding operations to use at runtime. 
We set the ranks of the communication group for synchronization.
Second, the data dependency of instructions can be added.  
Besides, the data dependency of instructions is generally related to stages. For instance, synchronization among submodules must be scheduled after the forward passes and before the backward passes corresponding to the final stages of each submodule in multi-modal models discussed above (Lines 14-19 of Figure~\ref{fig-dslmultimodal}). Therefore, a new stage must be registered where the new registered instruction is attached through \textit{register\_new\_stage} (Lines 12,13 of Figure~\ref{fig-dslmultimodal}). The data dependency of instructions is set through the method \textit{set\_cssr\_deps} (Line 19 of Figure~\ref{fig-dslmultimodal}). The set consisting of pairs $((inst\_type1,stage_i),(inst\_type2,stage_j))$ needs to be passed, which means the $inst\_type1$ corresponding to $stage_i$ is scheduled before the $inst\_type2$ corresponding to $stage_j$. Third, \textit{map\_inst\_to\_operation} maps a user-defined callable operation to the registered instructions at runtime (Line 20 of Figure~\ref{fig-dslmultimodal}). 
% We will elaborate on how the registered instructions influence the scheduler in \S~\ref{sec:cssr}.

% \begin{figure}[h]
% \centerline{\includegraphics[width=\linewidth]{figs/instapi.pdf}}
% \caption{APIs for registering new instructions.}
% \label{fig-newinsts}
% \end{figure}

\phb{Controllability in scheduling micro-batch computations.} We further provide multiple methods to control the scheduling order of micro-batch computations besides the scheduling priorities discussed above. Figure~\ref{fig-dslmicros} gives examples. 
% The parameters with the type \textit{callable} require users to write a function to fulfill their demands.
\textit{Config\_inflight\_micros} controls the schedule's maximum number of in-flight micro-batches. Users can specify in-flight micro-batches for each stage via a list through the parameter ``inflight-micros''. Besides, Users can also set rules by defining callable functions (Lines 1-6 of Figure~\ref{fig-dslmicros}(i)), which will be invoked by the scheduler (\S~\ref{sec:lib}) automatically. \textit{Register\_new\_checkfunc} sets rules during the scheduling of micro-batch computations by the scheduler (Line 16 of Figure~\ref{fig-dslmicros}(ii)). The example (Lines 1-15 of Figure~\ref{fig-dslmicros}(ii)) checks whether the instruction of type \textit{insttype} on actor \textit{pipeid} to be scheduled meets the data dependency. The corresponding instruction is to be scheduled if \textit{true} is returned.
\textit{Register\_new\_priority} enables users to register new priorities when the built-in values in Table~\ref{tab:compprimitives} for the scheduling priorities can not express the user-defined schedules (Line 3 of Figure~\ref{fig-dslmicros}(iii)). We will detail how the \textit{step\_func} is defined in \S~\ref{sec:aws}.
Users can invoke one or more methods according to their needs. 

\section{Scheduler}\label{sec:lib}
% This section presents the FlexPipe scheduler.
We break down the scheduling process into three phases as shown in Figure~\ref{fig-framework}. First, the model is partitioned into stages and then distributed over actors based on the stage placement strategy. Besides, the stages' DAG is maintained to track data dependencies.
This phase is executed once the stage data is specified as mentioned in Section~\ref{sec:dsl-data}.
Second, the computation schedule space representation (CSSR) is built. Third, the actor-aware schedule is performed to arrange schedules based on the scheduling settings concerning micro-batch computations described in Section~\ref{sec:dsl-schd}.
We leverage the optimizations on multi-modal models as an example to illustrate the scheduling process in Figure~\ref{fig-schedprocess}.

\begin{table}[t]
    \centering
    \caption{Built-in values for the stage placement strategy and the scheduling priorities.}
    \resizebox{\linewidth}{!}{
	\renewcommand\arraystretch{1.6}
    \begin{tabular}{ccc}
    \hline
       \multicolumn{2}{c}{\textbf{Name}} &  \textbf{Values} \\
      \hline
      \multicolumn{2}{c}{\textbf{Stage Placement Strategy}} & \makecell[l]{one-to-one, circular, \\ V-shape, bidirectional} \\
      \hline
      \multirow{2}{*}{\textbf{\makecell[l]{Controllability \\ in the Order of \\ Micro-batch\\ Computations \\ Over Stages}}} & \textbf{\makecell[l]{Computation Type \\ Traversal Priority}}  & \makecell[l]{(bwdfirst, unit1, unit2), \\ (fwdfirst, unit1, unit2), \\ (interleaved, unit1, unit2)} \\
      \cline{2-3}
                                                & \textbf{Stage Traversal Priority} & \makecell[l]{breadth-first, \\ (breadth-first, interval), \\ depth-first, \\ (depth-first, interval)} \\
    \hline
    \end{tabular}
    }
    \label{tab:compprimitives}
\end{table}
% \section{Scheduler}
% Stage placement strategies are critical to the performance of pipeline parallelism. 

\subsection{Computation Schedule Space}\label{sec:cssr}
We use the computation schedule space representation (CSSR) to enable various pipeline schedules. 
As is known, once the stage placement strategy is specified, each actor's scheduled micro-batch computations are determined. Each actor must constantly select the micro-batch computations whose data dependency has been resolved for scheduling until all assigned computations have been scheduled. 
Hence, CSSR is used to dynamically acquire the dependency-free micro-batch computations for each actor at each scheduling step.

% As shown in Figure~\ref{fig-overview}, CSSR is built according to the DAG of the stages and the actor data after applying the stage placement strategy.
% CSSR consists of all computation tasks for actors and the corresponding data dependencies. It also includes actor-aware scheduling states and dynamic data dependency derivation to support automated schedule arrangements.
% In CSSR, we use an instruction item to represent a micro-batch computation or communication operation. 
The main components of CSSR are listed below.

% \begin{itemize}
    \phb{Instruction Item.}
    % The task item is the basic task unit to be scheduled for each micro-batch over the stages.
    Instruction items are instruction instances used to represent the micro-batch computation or communication. 
    An instruction item is regarded as the basic unit scheduled in the scheduler.
    % to represent the corresponding micro-batch computation or communication operations.
    % The instruction items are categorized into multiple types listed in Table~\ref{tab:insts}, and 
    An instruction item is labeled with (instruction type, micro-batch ID, stage ID). 
    % The task item will be mapped to functions for execution at runtime.
    
    \phb{Instruction pool.} The instruction pool holds all the instruction items of all actors
    to be scheduled. Multiple instruction queues are created. Each instruction queue contains instruction items corresponding to all micro-batch computations over all stages of a specific instruction type. Instruction items are stored in the queues by micro-batch ID and stage ID. Instruction queues corresponding to the instruction type \textit{FwdPass} and \textit{BwdPass} are created by default. Besides, instruction queues corresponding to instruction types that users register using the scheduling language are also created.
    
    % Containing the task items representing forward and backward passes of different micro-batches over all stages.
    % The task pool also creates items representing undefined computations and allows users to implement custom operations at runtime.
    % Data dependency maintains the scheduling order of instruction items, which preserves the semantics of the DNN model. 
    
    \phb{Actor-aware reorder queues.} Reorder queues are used to maintain instruction items whose data dependencies have been resolved for actors. 
    Hence, each actor creates multiple reorder queues that correspond to different instruction types. By default, on an actor, reorder queues corresponding to \textit{FwdPass}, \textit{BwdPass}, and instruction types registered by users are created. Each reorder queue manages dependency-free instruction items corresponding to all micro-batch computations over the stages assigned to the actor to which the reorder queue belongs.

    \phb{Local buffer.} We configure a local buffer for each actor to keep the scheduling states of a scheduling step that mainly tracks the scheduled instruction items. Before the next scheduling step begins, local buffers are cleared. This is primarily intended to simulate a distributed environment in which the scheduling states are inaccessible to remote actors in the same scheduling step.

    \phb{Data dependency.} The scheduling of instruction items adheres to data dependencies, which means scheduling prior dependent instruction items is a prerequisite for scheduling subsequent instruction items. CSSR employs constructs to record data dependencies between instruction items. The data dependencies between instruction items for forward passes are transformed from the DAG of stages, the reverse of which are the data dependencies between instruction items for backward passes. Data dependencies set by \textit{set\_cssr\_deps} are also built. Besides, the instruction item with a smaller micro-batch ID is preferentially scheduled over a larger one of the same instruction type and stage ID by default.

    \phb{Dynamic data dependency resolver.} The data dependency resolver is leveraged to maintain the actor-aware reorder queues dynamically.
    % based on data dependencies between instruction items. 
    Once an instruction item is scheduled, the resolver immediately checks if its successors' data dependencies are resolved based on the scheduling states in the corresponding local buffer. Subsequent instruction items with resolved data dependencies are put into the reorder queues of the corresponding actor. 
% If we set priority for the traversal to the reorder queues of each actor, the prior actor fetches tasks regularly, and the tasks whose data dependencies have been resolved can be put into the reorder queues regularly for subsequent actors. 
% 3) The repetitive cycles require pre-processing to launch enough active micro-batch computations in the warm-up phase and post-processing in the cool-down phase. Besides, the warm-up and cool-down phases also use the same priority in traversal to the reorder queues.
% \# \textit{states}: tracking the scheduling states. \\

\begin{algorithm}[t]
\caption{StepFunction}\label{alg-sched}
\KwIn{\textit{actorid, states, cttp, stp, cfuncs}}
\KwResult{An instruction item or a nop.}
    ordered\_inst\_type $\leftarrow$ sort\_inst\_types(cttp) \\
    \For {insttype in ordered\_inst\_type}{
        queue $\leftarrow$ get\_reorder\_queue(actorid, insttype) \\
        direction, interval $\leftarrow$ parse\_stp(stp, insttype) \\
        \If{is\_empty(queue)}{
        continue
        }
        \If{\textit{interval} $\neq$ None} {
            pos, stepped\_interval $\leftarrow$ \textit{states} \\
            \If{stepped\_interval == interval}{
            pos $\leftarrow$ move(queue, pos, direction)\\
            }
            stageid, microid $\leftarrow$ fetch\_inst1(queue, pos, \textit{cfuncs}) \\
            return (insttype, stageid, microid) \\
        }
        \Else {
        stageid, microid $\leftarrow$ fetch\_inst2(queue, direction, \textit{cfuncs}) \\
        return (insttype, stageid, microid) \\
        }
    }
\end{algorithm}

\subsection{Actor-aware schedule}\label{sec:aws}
We use an actor-aware schedule mechanism that interacts with CSSR to arrange various schedules automatically. 

\textit{Key Insights.} With CSSR, each actor can constantly fetch instruction items for scheduling from the corresponding reorder queues. The correctness of the schedule is ensured through the dynamic data dependency resolver. The key to achieving high efficiency is determining which instruction items to schedule when multiple alternatives are available.

Furthermore, we observe that efficient pipeline schedules generally exhibit repetitive cycles, wherein the same computation types are periodically scheduled over different micro-batches on each actor.
The repetitive cycles can be maintained if each actor traverses reorder queues to fetch available instruction items for scheduling in a fixed order.
Various schedules can be arranged by enumerating different traversing orders along reorder queues.

Therefore, we translate the controllability in the scheduling of micro-batch computations into controlling the traversing order of instruction items in the reorder queues of each actor. We abstract two types of priorities to determine various traversing orders of reorder queues as detailed below.
% We set actor-aware priorities for the traversal of tasks to be scheduled.
% We also configure maximum active micro-batches for fewer bubbles.
% Once the stage placement is settled, the computation sequence order needs to be arranged for fewer pipeline bubbles. The scheduling priority primitives are leveraged to select the schedulable computation with the specific type (i.e. a forward pass or a backward pass) and the micro-batch of the mapped stages for each device in a scheduling step.
%  % and the micro-batch of the computation to be scheduled in a certain scheduling step, respectively. 
%  We design four kinds of parameterized scheduling priority primitives 
%  as shown in Table~\ref{tab:compprimitives} with which the mainstream PP methods can be expressed succinctly.
% In addition, the primitive \textit{WeakDependencyDesp} is also designed to apply user-defined scheduling priority to control the computation sequence order precisely. We will detail how to configure the primitives in the following.

 \phb{Computation Type Traversal Priority.} The computation type traversal priority determines 
 which type of computations (i.e., forward or backward passes) 
 to be prioritized for scheduling. 
 More specifically, it depicts which reorder queue (i.e., the one corresponding to \textit{FwdPass} or \textit{BwdPass}) to be traversed preferentially. 
 \textit{CompTypeTraversal} (Line 6 of Figure~\ref{fig-1f1bexp}) allow users to set different priorities.
 Besides, we build in three kinds of priorities (listed in Table~\ref{tab:compprimitives}), namely \textit{bwdpass-first, fwdpass-first}, and \textit{interleaved}, which refers to preferentially traversing the reorder queues corresponding to \textit{BwdPass}, \textit{FwdPass}, and interlaced \textit{BwdPass} and \textit{FwdPass}, respectively.
 % The computation type scheduling priority can be set through the primitive.
 We can also set different scheduling units $unit1$ and $unit2$ for \textit{FwdPass} and \textit{BwdPass} instruction items, respectively. By default, $unit1$ and $unit2$ are set to 1.

\begin{figure}[t]
\centerline{\includegraphics[width=\linewidth]{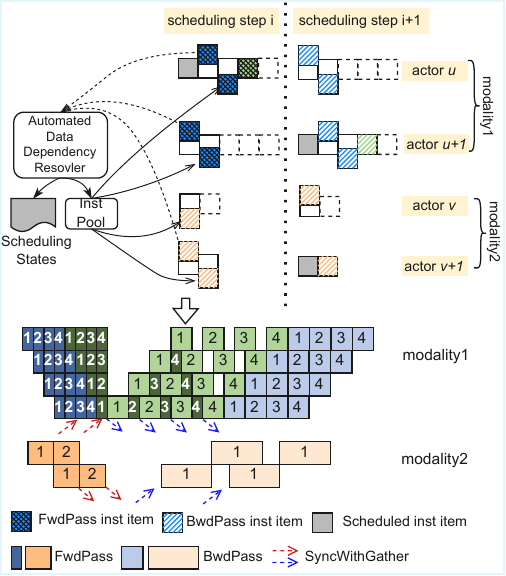}}
\caption{The automated scheduling process. The circular and one-to-one stage placement strategies are applied to ``modality1'' and ``modality2'', respectively. \textit{cttp=``interleaved'', fstp=(``breadth-first'',4), bstp=(``depth-first'',4))} for ``modality1''. \textit{cttp=``bwdpass-first'', fstp=breadth-first'', bstp=``breadth-first''} for ``modality2''.}
\label{fig-schedprocess}
\end{figure}

\phb{Stage Traversal Priority.} The stage traversal priority determines the micro-batch computations over which stage to be prioritized for scheduling when multiple stages
are placed on an actor. Concretely, it determines how to traverse the reorder queue corresponding to a specific instruction type.
% We suppose the micro-batches over the same stage are scheduled consistently with the micro-batch order, which does not affect performance~\cite{lin2024tessel}. 
% We need to determine the traversal scheduling priority for schedulable micro-batch computations of the mapped stages. 

We use \textit{StageTraversal} and \textit{interval} (Line 7 of Figure~\ref{fig-1f1bexp}) to control the direction and the speed of the traversal to a reorder queue, respectively.
We build in two types of traversal directions for \textit{StageTraversal}, including \textit{breadth-first} and \textit{depth-first}. \textit{Breadth-first} traverses the reorder queues from front to end and preferentially schedules instruction items concerning micro-batch computations of stages with model layers in the front end, while \textit{depth-first} traverses in the opposite direction and preferentially schedules instruction items concerning micro-batch computations of stages with latter model layers. Besides, when \textit{interval} is set with an integer, the current traversed instruction items with the same stage ID must be scheduled consecutively \textit{interval} times before the traversal moves forward to the instruction items with a different stage ID in the configured traversal direction. Otherwise, the first traversed instruction item in the traversal direction is scheduled.
Besides, the stage traversal priority can be set differently for \textit{FwdPass} and \textit{BwdPass} reorder queues.

% \phb{Configuration of In-flight Micro-batches.}
% The number of in-flight micro-batches
% influences the peak memory consumption and the order of micro-batch computations in pipeline schedules~\cite{narayanan2019pipedream}.
% We use \textit{config\_inflight\_micros} to configure the maximum number of in-flight micro-batches to limit forward pass scheduling. 
% By default, we schedule as many forward passes as possible under the memory limit in the warm-up phase~\cite{qi2023zero}.

After configuring priorities, actors schedule instruction items in the following steps.

\textit{Step 1.} Initialize reorder queues. Only \textit{FwdPass} instruction items of the first pipeline stage are dependency-free and
schedulable, and are put into the reorder queue corresponding to \textit{FwdPass} of the actor with the first stage.

\textit{Step 2.} In a scheduling step, all actors perform StepFunction in Algorithm~\ref{alg-sched} to fetch schedulable instruction items. If there are no schedulable alternatives, a bubble comes up.

\textit{Step 3.} Each actor updates the scheduling states in its local buffer.

\textit{Step 4.} The dynamic data dependency resolver checks the subsequent instruction items whose data dependency has been resolved and puts the schedulable alternatives into the corresponding reorder queues.

\textit{Step 5.} Update the scheduling states from the local buffer to the instruction pool and clear the local buffer. If all instruction items of the instruction pool have been scheduled, the scheduling process exits. Otherwise, return to \textit{step 2}.

Algorithm~\ref{alg-sched} presents how to fetch instruction items according to the built-in values of priorities.
Besides, the rules specified by users (Line 7/Line 16 of Figure~\ref{fig-dslmicros}(i)/(ii)) are checked through \textit{cfuncs} when fetching instruction items (Line 14 and 18 of Algorithm~\ref{alg-sched}.
% The StepFunction is defined according to the built-in values of the two priorities. 
Users can define new traversal orders of instruction items in reorder queues by writing a new callable step function and registering it as a new value for the two priorities through Line 3 of Figure~\ref{fig-dslmicros}(iii). 

\begin{figure}[h]
\centerline{\includegraphics[width=1.0\linewidth]{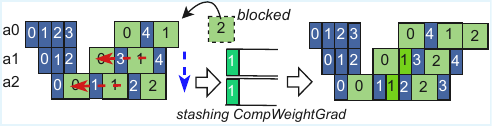}}
\caption{An example for the gradient separation algorithm.}
\label{fig-grad}
\end{figure}

\phb{Bubble optimizations.} We design and implement a gradient separation algorithm to automatically optimize bubbles of different schedules. As shown in Figure~\ref{fig-grad}, when a bubble is detected, we first estimate the blocked instruction item, e.g., a \textit{BwdPass} whose micro-batch ID is 2 on actor 0. Then, we backtrace the scheduled \textit{BwdPass} to stash weight gradient computations of actors 1 and 2 to schedule instruction items in the dependency chain as early as possible. With dependent instruction items scheduled, the blocked instruction items can be released to eliminate bubbles. The stashed weight gradients are inserted into the subsequent bubbles on the corresponding actors.
We employ asynchronous P2P communication that can be referred to in the work~\cite{jiang2024dynapipe}, and details are omitted here.
% It is worth noting that we do not need to make special arrangements for the warm-up and cool-down phases since the two phases fetch tasks in the same way as the steady phase.

\section{Auto-Tuner}\label{sec:autotune}
We use grid search to find optimal schedules in the parameter space defined by:
$$
(pp, dp, mbs, \underbrace{placement \ strategies, fstp, bfstp, ctp}_{schedule \ types}).
$$
The parameter values are set according to the following rules.

\begin{itemize}
\item The parameters determining the schedule types are taken from Table~\ref{tab:compprimitives}.
    \item The $interval$ is not set unless the circular placement is applied.
    \item The $ctp$ is not set to \textit{forward-first} unless the user specifies.
    \item $unit1$ and $unit2$ is set to 1 since we generally apply fine-grained gradient computation to reduce bubbles.
\end{itemize}

\begin{figure*}[t]
\centerline{\includegraphics[width=\linewidth]{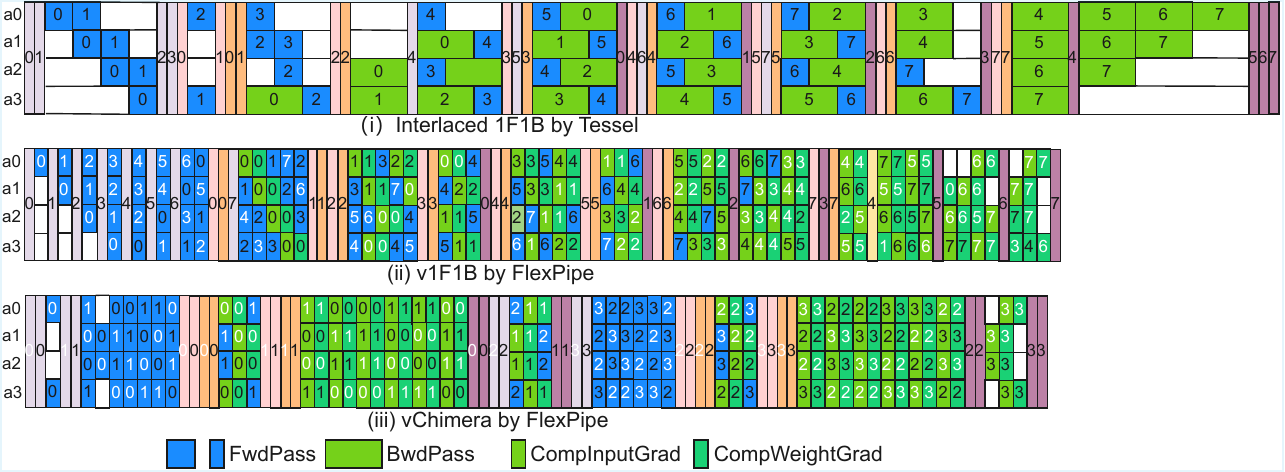}}
\caption{Schedules explored by Tessel and FlexPipe.}
\label{fig-expsched}
\end{figure*}

% \phb{Performance Estimator}
% We draw on the idea of Proteus, which is a simulator for distributed parallelism. 
% Besides, it provides a set of PyTorch-like interfaces that infer the shapes of inputs and outputs of different layers given a model definition without instantiating tensor data. However, it has two drawbacks. First, the shape inference for a whole model is too
% time-consuming. Besides, it only supports 1F1B.

% We use the performance estimator to predict the performance of the alternatives during the tuning. First, the schedule is transformed into the execution graph of tasks, which are labeled with (start time, end time, cost). The profiler is used to acquire the cost of independent layers, mainly including the Embedding layer and the transformer layer. Suppose we use $t^l$ and $t^s$ to represent the cost of a layer and a stage, then,
% $$
% t_s = \sum_{i=u}^{v}{t^l_i}.
% $$

\section{Evaluation}
% We evaluate FlexPipe for two aspects. First, we present the searched schedules when a stage placement strategy is specified.
% Second, we verify the programmability of FlexPipe by employing different schedules and registering asynchronous synchronization
% operations based on the concise interfaces in Section~\ref{sec:dsl} for the optimization of multimodal models.
% Furthermore, pipeline and data parallelism are used to scale up to more nodes in all experiments.
% The end-to-end training performance of all schedules and optimizations is presented and compared with state-of-the-art works. We also conduct ablation studies to analyze the detailed performance of different schedules.

We conduct experiments on up to 64 NVIDIA A800 SXM4 80GB GPUs distributed across 8 nodes.

\subsection{Automated Exploration of Schedules.}
In this section, FlexPipe automatically searches efficient schedules for transformer models with large embedding layers. Two main model configurations with varying vocabulary sizes are tested as shown in Table~\ref{tab:gptcfg}. 

\begin{table}[t]
    \centering
    \caption{Model configurations of GPT models.}
    \resizebox{\linewidth}{!}{
	\renewcommand\arraystretch{1.3}
    \begin{tabular}{ccc}
      \hline
      \makecell[c]{Model Size without \\ Embedding Layers} & 5B & 16.1B\\
      \hline
      Total Model Size & 5.4B/5.7B/6.3B/7.7B/10.2B & 18.2B/20.3B/24.3B/25.1B/25.9B\\
      \hline
      Vocabulary Size & 64K/128K/256K/512K/1M & 256K/512K/1M/1.1M/1.2M\\
      \hline
      Layers & 64 & 80\\
      \hline
      Attention Heads & 2560 & 4096\\
      \hline 
      Hidden Size & 64 & 32 \\
      \hline
      Sequence Length & 512 & 1024\\
      \hline
      Global Batch Size & 128 & 128 \\
      \hline
    \end{tabular}
    }
    \label{tab:gptcfg}
\end{table}

\begin{table}[t]
    \centering
    \caption{Model configurations of CLIP models.}
    \resizebox{0.8\linewidth}{!}{
	\renewcommand\arraystretch{1.1}
    \begin{tabular}{ccccccc}
      \hline
     Total Model Size & \multicolumn{2}{c}{4.7B} & \multicolumn{2}{c}{8.2B} & \multicolumn{2}{c}{9.7B} \\
      \hline
     Submodule & Audio & Text & Video & Text & Image & Text \\
     \hline
     Submodule Size & 3B & 1.7B & 5.5B & 2.7B & 6.5B & 3.2B \\
      \hline
      Layers & 240 & 240 & 440 & 216 & 32 & 40 \\
      \hline
      Attention Heads & 16 & 16 & 16 & 16 & 64 & 40 \\
      \hline
      Hidden Size & 1024 & 768 & 1024 & 1024 & 4096 & 2560 \\
      \hline
      Sequence Length & 264 & 77 & 264 & 77 & 264 & 77 \\
      \hline
      Global Batch Size & \multicolumn{2}{c}{128} & \multicolumn{2}{c}{128} & \multicolumn{2}{c}{128} \\
      \hline
    \end{tabular}
    }
    \label{tab:instructions}
\end{table}

\phb{Baselines.} We compare FlexPipe with three baselines employing different schedules: 1)Megatron-LM, which uses 1F1B; 2) T-1F1B, which adapts 1F1B to incorporate the MLLM stage placement strategy in Figure~\ref{fig-placements}, and the corresponding schedule of T-1F1B is arranged by FlexPipe; 3) Tessel, which automatically searches schedules for the MLLM stage placement strategy. In the MLLM stage placement strategy, tensor parallelism of embedding layers is applied along the vocabulary dimension and distributed to the GPUs within the same pipeline parallelism group.

We search the space of the parameters $(pp, dp, mbs)$ (for power-of-two) to find the best performance for each baseline. For a specific parameter combination $(pp, mbs)$, Tessel automatically searches pipeline schedules. We illustrate the schedule searched by Tessel when $pp=4$ in Figure~\ref{fig-expsched}.

FlexPipe searches the space described in Section~\ref{sec:autotune}. Parameters $pp$, $dp$, and $mbs$ are enumerated in the same way as those of the baseline. Given a parameter combination $(pp,mbs)$, FlexPipe automatically searches schedules and returns the alternatives with the least bubble ratio for different stage placement strategies. Specifically, the schedule space is defined by $(stage \ placement \ strategy,$ $fstp, bstp, ctp)$, each of which is enumerated from the corresponding built-in values in Table~\ref{tab:compprimitives}. A combination of V-shape and bidirectional stage placement strategies is also considered. Besides, we use tensor parallelism along the vocabulary dimension for each stage placement strategy to distribute the first and last layer over GPUs within the pipeline parallelism group. The \textit{interval} is set to $pp$ for the option \textit{(breadth-first, interval)} and \textit{(depth-first, interval)} of the stage traversal priority $fstp$ and $bstp$. We illustrate the schedule searched by FlexPipe when $pp=4$ in Figure~\ref{fig-expsched}. We evaluate the two types of schedules since the bubble ratios of the two are close. Besides, even though vChimera uses the bidirectional stage placement strategy, which shows fewer bubbles, its efficiency may degrade due to more synchronization between the last stages, especially with large GPUs. 

We compare the search cost of Tessel and FlexPipe in Table~\ref{tab:searchcost}, where the ``X'' mark represents that the search process fails. Tessel searches the repend, warm-up, and cool-down phase based on the Z3-solver. Besides, the search space shows explosive growth with the increased number of devices.  The results show that the search cost by Tessel is the poorest among the three search methods, and the search process fails when the number of devices reaches 8. 
To improve the search cost by Tessel, Tessel-fast uses several algorithms to construct schedules shown in Figure~\ref{fig-expsched}. However, the search still fails when the number of devices reaches 16.
FlexPipe uses the least search cost of the three search approaches. The search cost increases gently with the increased number of devices. It also searches for multiple efficient alternatives for different model configurations.

\begin{table}[h]
    \centering
    \caption{Comparison of search cost by FlexPipe and Tessel.}
    \resizebox{0.65\linewidth}{!}{
	\renewcommand\arraystretch{0.8}
    \begin{tabular}{cccc}
    \toprule
     GPUs & Tessel & Tessel-fast & FlexPipe \\
     \hline
     2 & 1.29s & 0.26s & 4.73s\\
     \hline
     4 & 163.62s& 11.1s & 8.28s\\
     \hline
     8 & X & 729.74s & 71.28s\\
     \hline
     16 & X & X & 107.34s\\
     \hline
     32 & X & X & 534.78s\\
      \bottomrule
    \end{tabular}
    }
    \label{tab:searchcost}
\end{table}

\begin{figure}[t]
\centerline{\includegraphics[width=0.85\linewidth]{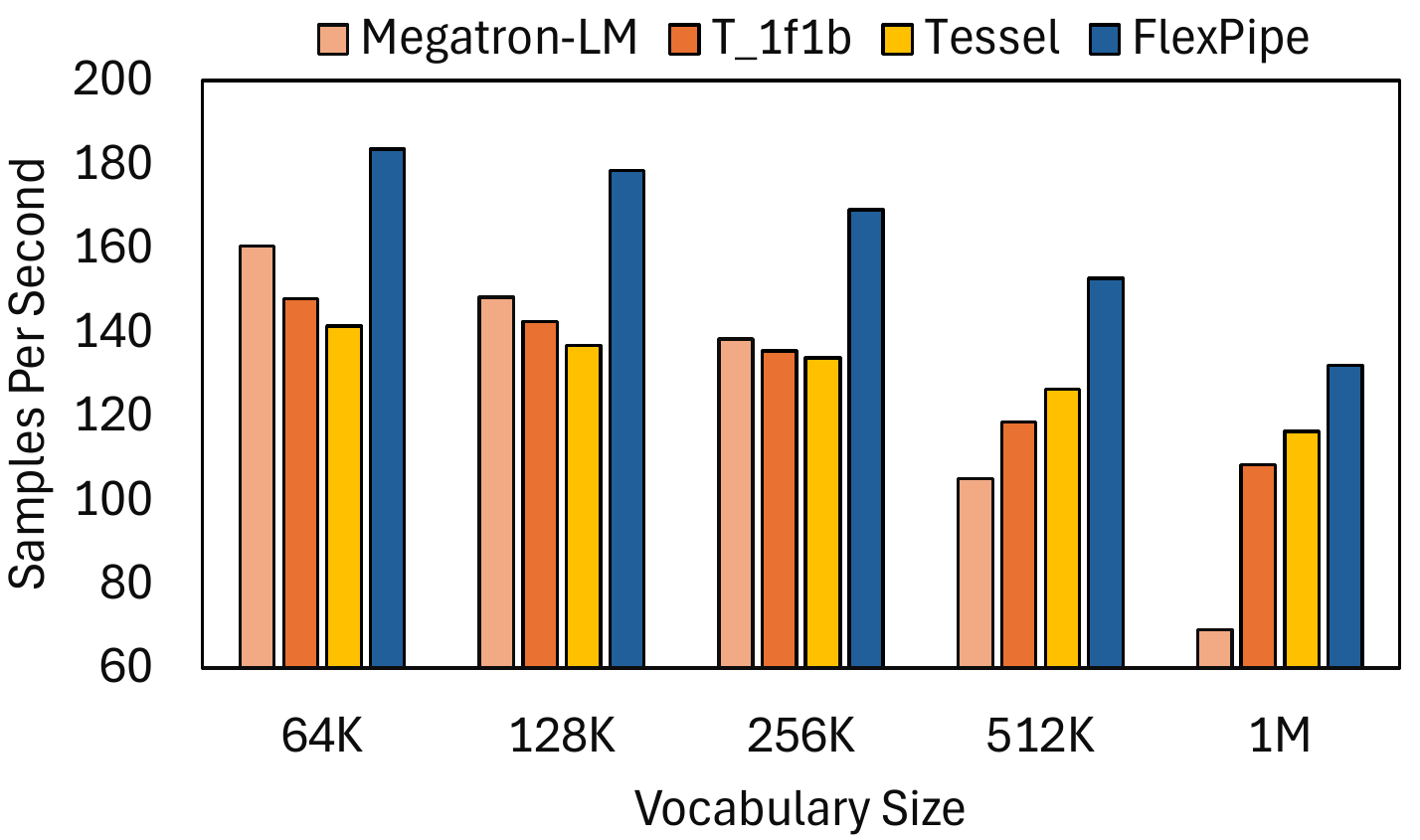}}
\caption{End-to-end performance of 5B GPT models.}
\label{fig-perf5b}
\end{figure}

\phb{End-to-end performance.} Figure~\ref{fig-perf5b} compares the end-to-end performance results of 5B GPT models with varying vocabulary sizes from 64K to 1M on 32 GPUs. The experimental results demonstrate that FlexPipe gains 1.14$\times$-1.91$\times$, 1.20$\times$-1.28$\times$, and 1.13$\times$-1.30$\times$ speedup compared with Megatron-LM, T-1F1B, and Tessel, respectively. The performance of Megatron-LM degrades due to the imbalanced workloads incurred by the large vocabulary size. The larger the vocabulary size, the more the performance declines. T-1F1B introduces bubbles because of data dependency between AllReduce operations due to the tensor parallelism of the Embedding layers. Tessel launches more in-flight micro-batches to avoid the bubbles similar to T-1F1B. However, more in-flight micro-batches are launched in the warm-up phase, introducing more bubbles in the warm-up and cool-down phase as shown in Figure~\ref{fig-expsched}. The schedules searched by FlexPipe show fewer bubbles and gain a prominent performance improvement compared to the schedules of the three baselines. FlexPipe utilizes a schedule space that ranges from a wider spectrum of schedule types. Besides, the searched alternatives are further optimized by gradient separation.

\begin{figure}[t]
\centerline{\includegraphics[width=0.85\linewidth]{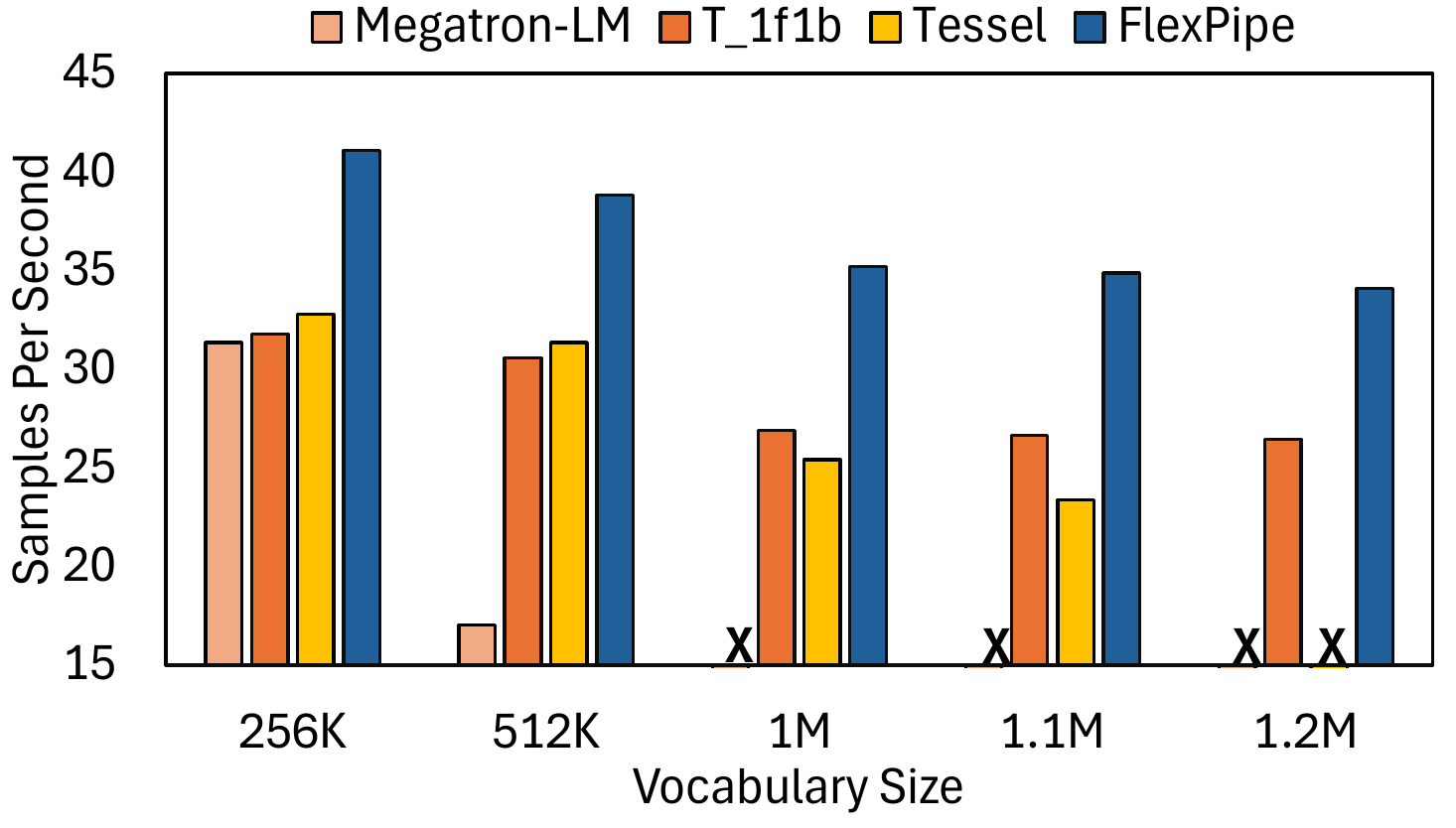}}
\caption{End-to-end performance of 16.1B GPT models.}
\label{fig-perf16b}
\end{figure}

We also compare the end-to-end performance results of 16.1B GPT models with varying vocabulary sizes from 256K to 1.2M on 32 GPUs in Figure~\ref{fig-perf16b}. The ``X'' mark represents that the out-of-memory issue is still encountered even if the recomputation is used. FlexPipe also shows superior performance to that of the other three baselines. More specifically, FlexPipe achieves 1.31$\times$-2.28$\times$, 1.27$\times$-1.31$\times$, and 1.24$\times$-1.49$\times$ speedup compared with Megatron-LM, T-1F1B, and Tessel, respectively. Megatron-LM requires more memory consumption due to the large vocabulary size, and encounters the out-of-memory issue when the vocabulary size reaches 1M. It is worth noting that $pp$ searches under 16 for Tessel since Tessel cannot support the automated schedule exploration when the number of GPUs reaches 16. Hence, the schedule by Tessel also encounters the out-of-memory issue when the vocabulary size reaches 1.2M.

\begin{figure}[h]
% \centerline{\includegraphics[width=0.5\linewidth]{figs/breakdownnew5b_cropped.pdf}}
\centering
\subfigure[5B GPT models]{\includegraphics[width=0.49\linewidth]{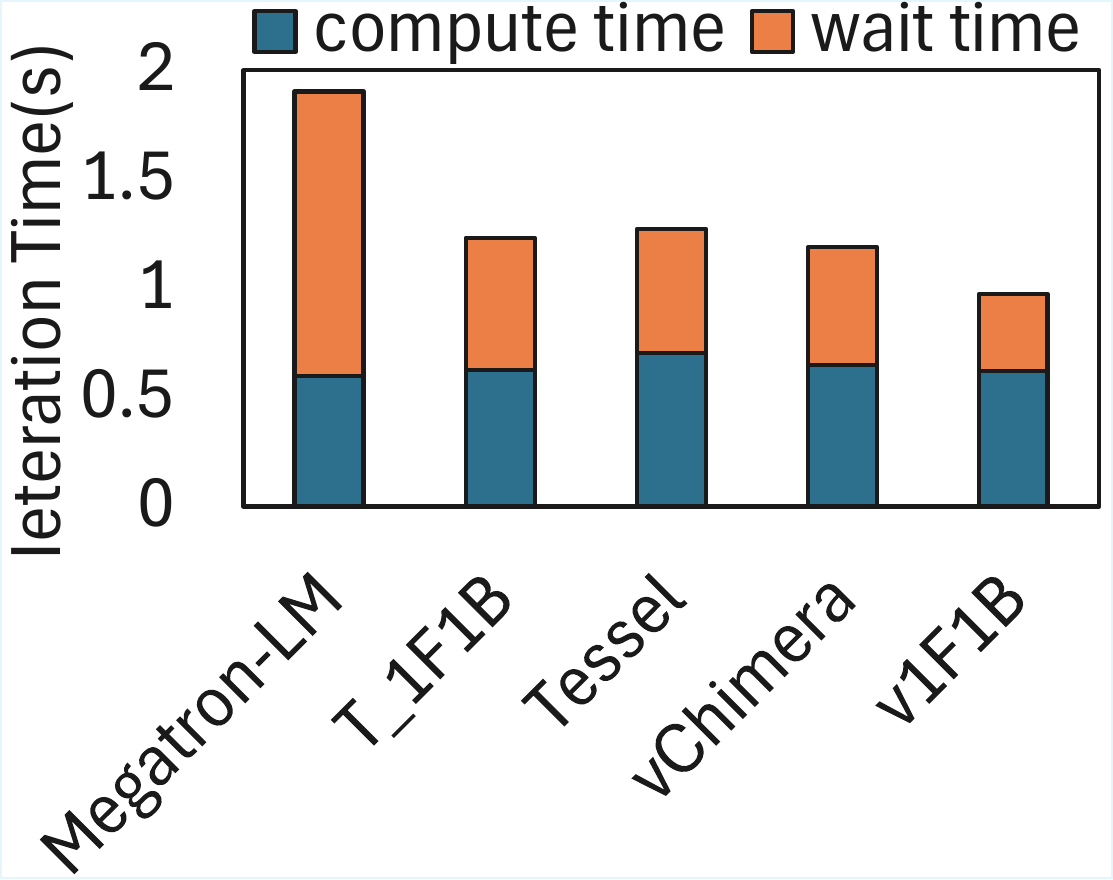}}
\subfigure[16.1B GPT models]{\includegraphics[width=0.49\linewidth]{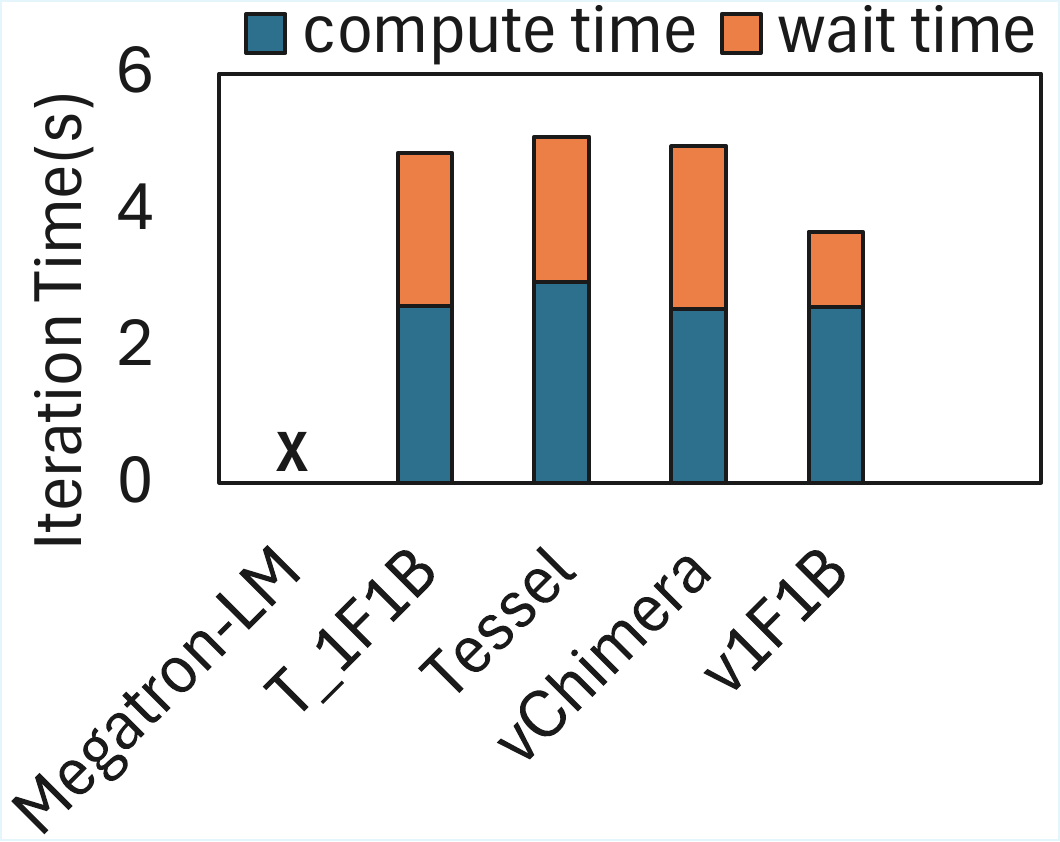}}
\caption{Runtime performance breakdown.}
\label{fig-breakdown5B}
\end{figure}

\phb{Performance breakdown.} Bubbles are the key factor that affects the efficiency of pipeline parallelism. Bubbles can be caused by data dependency and communication operations such as AllReduce by the tensor parallelism and P2P communication to exchange activations and gradients. To accurately determine bubbles in the schedules, we profile the runtime of an iteration for each schedule, break down the execution into the stage computation time and device waiting time, and use the device waiting time to reflect bubbles. Figure~\ref{fig-breakdown5B} presents the runtime breakdown of the 5B and 16.1B models with a 1M vocabulary size, respectively. The computation time of different schedules does not differ much, since each schedule must compute the same whole model. The slight difference lies in the computation efficiency caused by different micro-batch sizes. A larger micro-batch size generally has higher computation efficiency. The device waiting time exhibits a noticeable difference among different schedules, consistent with the end-to-end performance results.
Megatron-LM shows the largest device waiting time. The device waiting time of T-1F1B and Tessel is also larger than that of FlexPipe.

% \begin{figure}[h]
% \centerline{\includegraphics[width=0.5\linewidth]{figs/breakdownnew16b_cropped.pdf}}
% \caption{Runtime performance breakdown.}
% \label{fig-breakdown16B}
% \end{figure}

\begin{figure}[h]
\centerline{\includegraphics[width=1.0\linewidth]{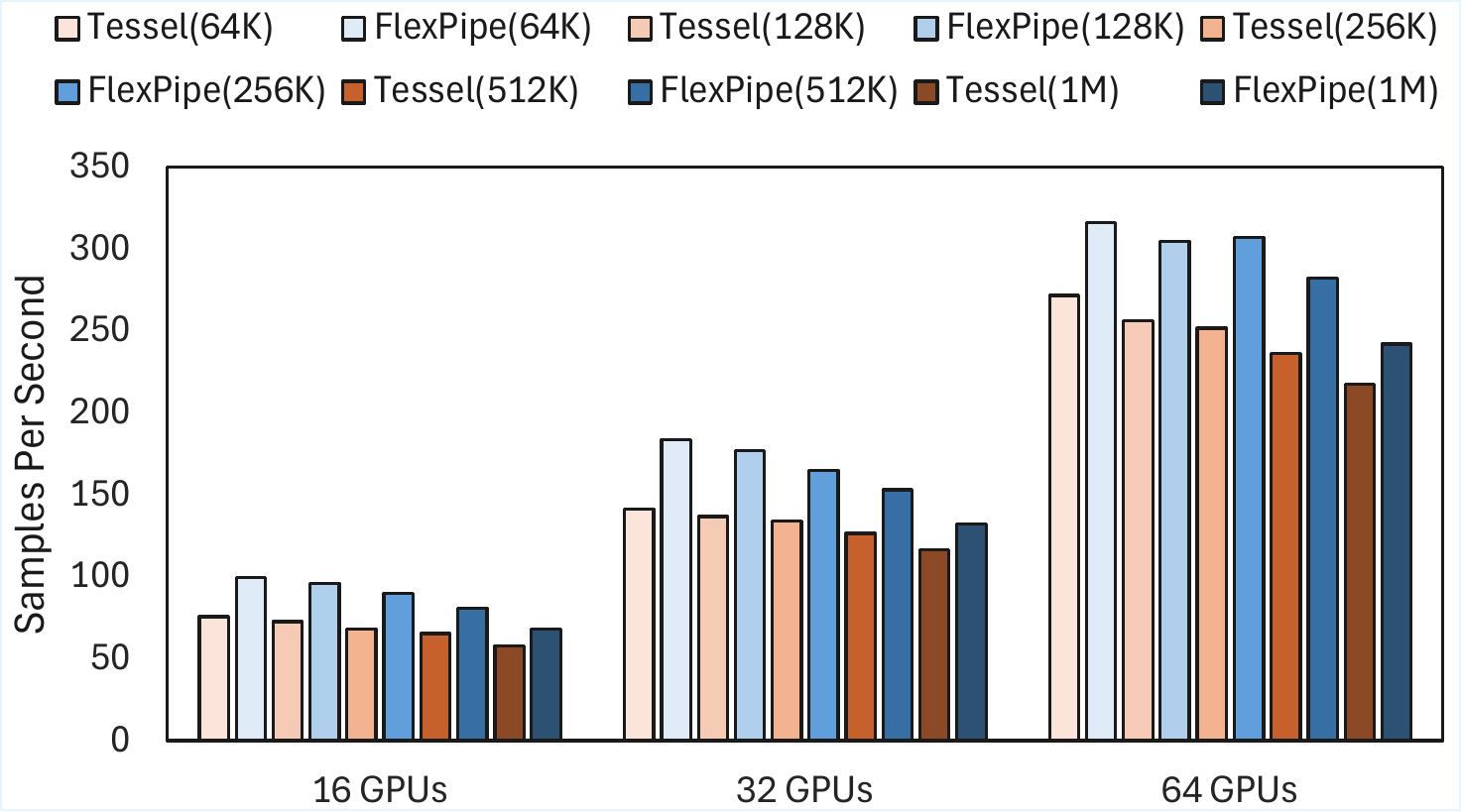}}
\caption{Weak scaling for 5B GPT models.}
\label{fig-scaling5b}
\end{figure}

\phb{Parallel scalability.} We present the weak scaling results on 5B GPT models in Figure~\ref{fig-scaling5b}. The number of GPUs increases from 16 to 64. Each schedule employs the same parameter configuration with which the schedule achieves optimal end-to-end performance results, but scales $dp$ with an exact multiple of the number of GPUs. The experimental result shows that FlexPipe achieves superior performance with different hardware resources.

% \begin{figure}[h]
% \centerline{\includegraphics[width=1.0\linewidth]{figs/scaling-16B.pdf}}
% \caption{FlexPipe Overview.}
% \label{fig-scaling16b}
% \end{figure}

\subsection{Optimizations on Multimodal Models}
% We optimize multimodal models based on FlexPipe. 

\begin{figure}[h]
\centerline{\includegraphics[width=0.83\linewidth]{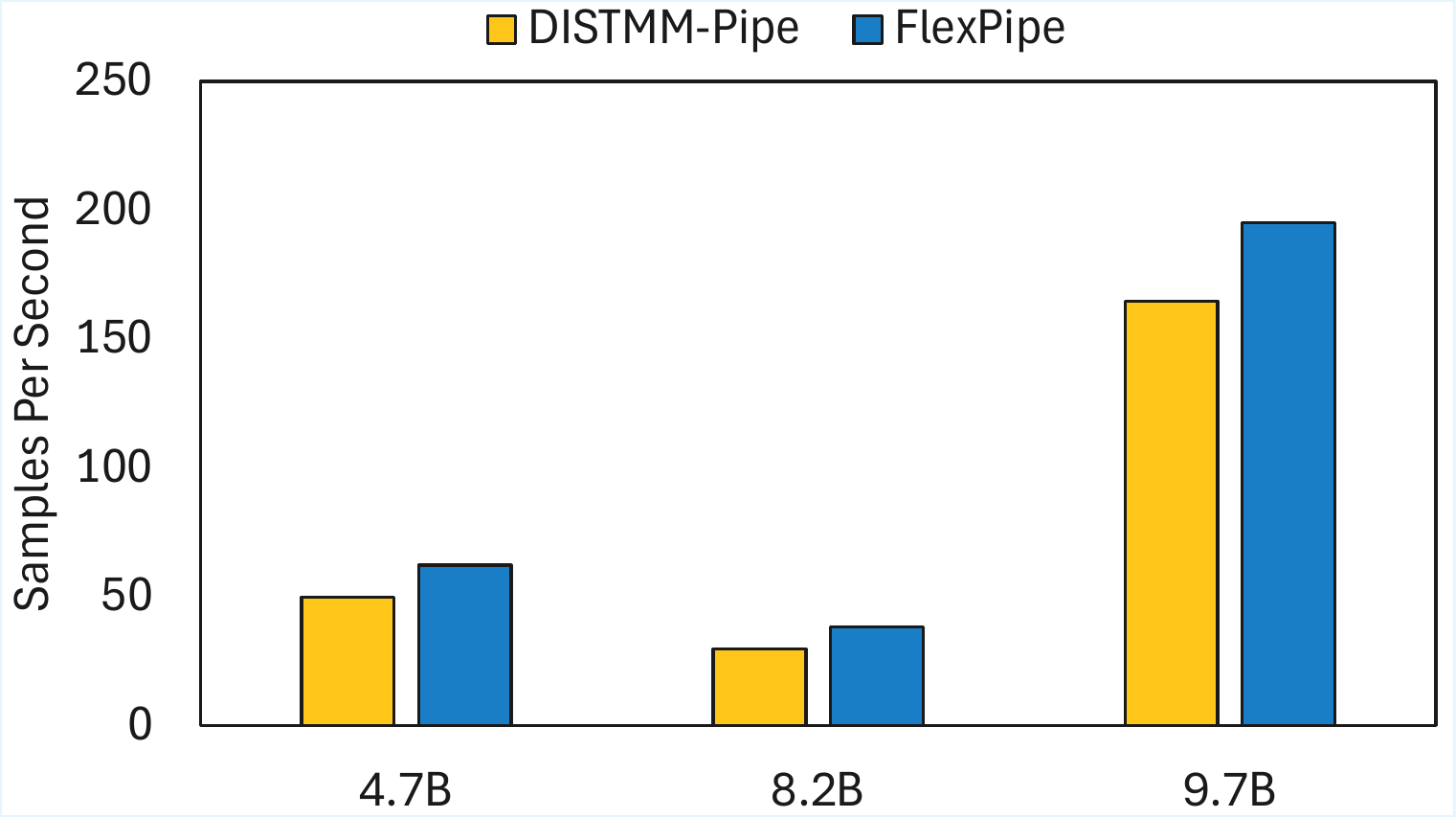}}
\caption{End-to-end performance of CLIP models.}
\label{fig-perfmultimodal}
\end{figure}

Compared with DistMM-Pipe, we use asynchronous communications to exchange the output states of submodules corresponding to different modalities. The asynchronous communication is registered as the new instruction \textit{SyncWithGather} as elaborated in Figure~\ref{fig-dslmultimodal}. The instruction \textit{SyncWithGather} is scheduled following each corresponding forward pass of the last pipeline stage for both modalities. Furthermore, we employ mixed schedules for different submodules to enhance efficiency. As in our experiments, we use interleaved 1F1B to calculate the audio/image modality, while applying 1F1B to the text modality. The schedule by FlexPipe is shown in Figure~\ref{fig-schedprocess}. Figure~\ref{fig-perfmultimodal} presents the end-to-end performance of CLIP models on 40 GPUs. Compared with DistMM-Pipe, FlexPipe gains 1.26$\times$, 1.29$\times$, and 1.18$\times$ performance speedup for CLIP models with size 4.7B, 8.2B, and 9.7B, respectively. The efficiency of the models with the former two sizes is comparatively low due to the small hidden sizes and attention heads. 
% Furthermore, we profile DistMM-Pipe and find that above 10\% cost is caused by synchronization, while our methods reduce synchronization cost. 
% Besides, the gradient separation can achieve up to 7\% performance gain.

\section{Related Works}
\phb{Schedules and Optimizations.} Exsiting research has primarily focused on handcrafting efficient schedules for performance. GPipe~\cite{huang2019gpipe} leverages batch splitting to enable devices to work simultaneously. 1F1B~\cite{narayanan2019pipedream} proposes early backward scheduling to reduce activation memory consumption. Subsequent schedules~\cite{narayanan2021efficient,li2021chimera,liu2023hanayo,qi2023zero,lamy2023breadth,wu2024bitpipe} have built upon the two techniques to reduce bubbles further. Multiple works~\cite{athlur2022VarunaScalable,sun2024adapipe,liu2025mario,huang2025obscura} leverage recomputations in pipeline parallelism methods to reduce memory consumption. Optimizations on long sequence scenarios are also proposed~\cite{lin2025weipipe}.
% Since PP generally enables multiple active micro-batches for fewer pipeline bubbles, bulky memory needs to be allocated to store activations for backward propagation. 
% Recomputation~\cite{chen2016training} and host memory offloading~\cite{zhou2023mpress} are effective methods to reduce activation memory consumption. 
% Varuna~\cite{athlur2022VarunaScalable} combines pipeline parallelism and activation recomputation, which designs a static rule-based schedule based on constraints including activation recomputation timing, activation memory management, and backward computation prioritization for computation and memory efficiency optimizations.
% Besides, Several PP methods, such as 1F1B, generally have imbalanced memory consumption due to early backward scheduling. Pipeline schedules with bidirectional or circular stage placements usually are out of the imbalanced memory issue since each device's computations are evenly scheduled.
Bpipe~\cite{kim2023bpipe} and Mpress~\cite{zhou2023mpress} 
% transfer activations from devices with larger memory costs to devices with spare memory to 
optimize the unbalanced memory requirements of different devices.
% AdaPipe~\cite{sun2024adapipe} designed a fine-grained recomputation scheme to improve the performance degradation due to imbalanced memory when recomputation is applied.
% mCAP~\cite{dreuning2022mcap} utilizes an
% incremental-profiling approach to partition models evenly
% across GPUs concerning peak memory usage. 
Our work supports most existing schedules and can further integrate various optimization approaches mentioned above by implementing optimization passes.

\phb{Frameworks.} 
DynaPipe~\cite{jiang2024dynapipe} uses a dynamic micro-batching approach to
the multi-task training of LLMs. Tessel~\cite{lin2024tessel} is a two-phase approach
% including repetitive pattern construction
% and schedule completion, 
to explore efficient schedules for various stage placement strategies automatically.
% DISTMM~\cite{huang2024distmm} launches doubled micro-batches to solve
% the dependency barrier caused by multi-modal training for computation efficiency.
GraphPipe~\cite{jeon2024graphpipe} uses a graph structure to partition DNN models to balance workloads. 
Our work covers a broader range of searchable schedule types than Tessel. Besides, our work can further use the partition algorithm by GraphPipe to improve efficiency.
% Koala~\cite{tang2025koala} proposes a two-structured DSL to develop and customize schedules.
% The FlexPipe DSL exhibits enhanced provivity and provides users with flexible controllability in micro-batch computations.

% with branch structures efficiently and execute stages with heavy workloads concurrently on multiple devices to
% improve pipeline utilization and reduce memory consumption.

\section{Conclusion}
FlexPipe explores efficient schedules automatically based on a succinct DSL and an automated scheduler with negligible overheads. It also provides programmable mechanisms to customize schedules for diverse model architectures swiftly. Evaluation results demonstrate prominent performance improvement compared with existing practices.
We look forward to extending FlexPipe with more optimization approaches to enhance efficiency for various scenarios.

\end{document}